\documentclass[aps,pra,showpacs,notitlepage]{revtex4-1}

\usepackage{graphicx}

\begin{document}

\title{Local measurement uncertainties impose a limit on non-local quantum correlations}

\author{Holger F. Hofmann}
\email{hofmann@hiroshima-u.ac.jp}
\affiliation{
Graduate School of Advanced Sciences of Matter, Hiroshima University,
Kagamiyama 1-3-1, Higashi Hiroshima 739-8530, Japan
}

\begin{abstract}
In quantum mechanics, joint measurements of non-commuting observables are only possible if a minimal unavoidable measurement uncertainty is accepted. On the other hand, correlations between non-commuting observables can exceed classical limits, as demonstrated by the violation of Bell's inequalities. Here, the relation between the uncertainty limited statistics of joint measurements and the limits on expectation values of possible input states is analyzed. It is shown that the experimentally observable statistics of joint measurements explain the uncertainty limits of local states, but result in less restrictive bounds when applied to identify the limits of non-local correlations between two separate quantum systems. A tight upper bound is obtained for the four correlations that appear in the violation of Bell's inequalities and the statistics of pure states saturating the bound is characterized. The results indicate that the limitations of quantum non-locality are a necessary consequence of the local features of joint measurements, suggesting the possibility that quantum non-locality could be explained in terms of the local characteristics of quantum statistics.
\end{abstract}


\maketitle

\section{Introduction}

Scientific investigations must be rooted in reproducible observations. In quantum theory, this requires a closer look at the mechanisms of measurement and control, since it is not immediately obvious how the theoretical formalism describes these mechanisms. In recent years, there has been a renewed interest in the non-classical aspects of measurement statistics, motivated to a great extent by an increasing variety of experimental realizations of quantum measurements \cite{Pry04,Bae08,Yok09,Ste09,Iin11,Gog11,Suz12,Roz12,Tan13,Bae13,Boy13,Rin14,Kan14,Den14,Bam14,Val16,Pia16,Iin16,Suz16,Pia17,The17,Iin18}. At the center of these investigations is the uncertainty principle and the associated possibility of jointly measuring complementary observables that do not have a joint representation in the theoretical description of quantum systems \cite{Eng96,And05,Hei08,Sta08,Yu10,Hei10,Car11,Lor11,Car12,Bus13,Hei13,Hof14,Kun14,Kin15,Kaw18}. The results of these investigations show that the Hilbert space formalism imposes a set of well defined rules on joint measurements, and the derivation of these rules from the structure of Hilbert space may help to explain the physics behind the strange and unexpected features of quantum theory. 

Of particular importance would be a better understanding of the violation of Bell's inequalities, which is widely recognized as the most compelling proof that the outcomes of quantum measurements cannot be explained in terms of local physical properties assigned to the individual systems \cite{Bell,Gen05,Car18}. It should be obvious that the violation of Bell's inequalities is only possible because there is no uncertainty free joint measurement of the observables concerned \cite{Wol09,Qui14,Uol14}, but it is less clear how non-local correlations appear in the statistics of uncertainty limited joint measurements. Up to now, the focus of the analysis has mostly been on formal criteria of non-classicality \cite{Ste14,Kar15,Qui16,Hir18}, and the results are usually related to the structure of Hilbert space algebras. The closest that previous research has come to a detailed analysis of the relation between measurement incompatibility and Bell's inequalities is the explanation of the Cirel'son bound \cite{Cir80}, which identifies the quantum mechanical limit of Bell's inequality violations. In the first analysis of this kind, Oppenheim and Wehner explained the Cirel'son bound using the uncertainty limits of conditional quantum states obtained from measurements of one of the two correlated systems \cite{Opp10,Car19}. Motivated by this result, it was then shown that the Cirel'son bound can also be derived from the uncertainty limits of a joint measurement of two collective observables in Bell's inequalities \cite{Ban13}. In a sense, these two approaches are complementary to each other, since \cite{Opp10} establishes the possibility of explaining the limit of non-local correlations in terms of the uncertainty limits of local quantum states without any analysis of joint measurability or measurement uncertainties, while \cite{Ban13} analyzes the limits of a collective measurement of the two qubits, thereby establishing a link between the Cirel'son bound and measurement incompatibilities in the four dimensional Hilbert space of two qubits. Specifically, \cite{Ban13} explains the Cirel'son bound as the uncertainty bound for a joint measurement of two incompatible correlations represented by $\hat{X}_A\hat{X}_B-\hat{Y}_A\hat{Y}_B$ and $\hat{X}_A\hat{Y}_B+\hat{Y}_A\hat{X}_B$. The joint measurement considered is therefore non-local and does not provide any information about the correlation between the local qubit components $\hat{X}_i$ and $\hat{Y}_i$. Since the concept of quantum non-locality relies on ambiguities in the relation between local quantum statistics and their non-local correlations, it should be of considerable interest to see whether there exists a relation between the Cirel'son bound and the uncertainties of local measurements performed separately on the two qubits. It is therefore the goal of this paper to explain the quantum mechanical bounds on non-local correlations in terms of the local measurement uncertainties of two joint measurements applied independently to each of the two quantum correlated systems. 

As I show in the following, the problem that needs to be solved first is the relation between quantum statistics and measurement uncertainties in a locally performed single qubit measurement. Although the mathematical formalism has been discussed at length in the literature, it seems that not sufficient attention has been paid to the experimentally observable correlations between the measurement outcomes for two non-commuting observables. I therefore start the analysis by pointing out that all uncertainty limited joint measurements of the qubit components $\hat{X}$ and $\hat{Y}$ result in outcome statistics where the average of the product of the two outcome values of $\pm 1$ is exactly zero. This experimentally verifiable fact explains the need for quantum state uncertainties, since it would result in the prediction of negative experimental probabilities for any combination of uncertainty free inputs and measurements. 
Importantly, the problem of negative probability predictions is not resolved by the measurement uncertainties alone. Once measurement uncertainties are included, the positivity of joint measurements defines a corresponding uncertainty bound that apply to the possible statistics of the input states. This means that the statistical rules that apply to joint measurements of two observables impose non-trivial limitations on the possible statistics observed in precise measurements of the individual observables. As a consequence, it is possible to derive the uncertainty bounds of local quantum states from the positivity bounds defined by the state-independent statistical properties of local joint measurements. 

The main result of the present paper is that the requirement of positivity for experimental  probabilities observed in a combination of two local joint measurements performed on two separate quantum systems defines a bound on the non-local correlations between the two systems (Eq.(\ref{eq:tightB}) below). It should be noted that this bound is much stronger than the Cirel'son bound, which only describes the upper limit of the violation of Bell's inequalities. The complete bound also restricts other linear combinations of the four correlations that appear in Bell's inequalities, producing a more precise description of the limits of non-local quantum coherence in entangled states. Bell's inequalities can be violated because the bounds imposed by a combination of two local measurement uncertainties is weaker than the bounds imposed by local realism. It is of particular importance that the result reported here explains the bounds on non-local correlations between two qubits in terms of local measurement uncertainties for joint measurements of single qubits. The limits of quantum correlations between two qubits can thus be traced back to the statistics of individual qubits, indicating that the failure of realism signified by Bell's inequality violations originates from a more fundamental failure of realism in the local relations between $\hat{X}_i$ and $\hat{Y}_i$ observed in joint measurements.  

\section{Statistics of joint measurements for qubits}
\label{Sec:measure}

Quantum theory sets very precise conditions for the realization of measurements that are simultaneously sensitive to two non-commuting observables. If the measurements are carried out in sequence, there is a necessary trade-off between the resolution of the first measurement and the disturbance of the second observable caused by this measurement. As a consequence, both measurement results are characterized by statistical errors, one caused by limited resolution and the other by a dynamical change of the system properties. However, the measurement statistics themselves do not distinguish between the possible causes of the errors or the sequence in which the measurements are performed, and the relation between input states and the joint probabilities of the measurement outcomes can always be summarized by assigning a measurement operator to each pair of outcomes. It is therefore possible to describe a set of universal conditions imposed by quantum theory on the joint measurement of any two non-commuting observables, independent of the details of the particular realization of the joint measurement. In general, these fundamental characteristics of joint measurements describe both the uncertainty limits and the experimentally observed correlations between the two measurement results in terms of a set of rules that describe the precise relation between the probabilities of measurement outcomes and the statistics of the input state. 

In the case of a single two level system, the rules that determine the experimentally observable statistics of measurement outcomes are particularly simple. For a pair of observables $\hat{X}$ and $\hat{Y}$ with eigenvalues of $\pm 1$ and mutually unbiased eigenstates, the quantum statistics of the input state can be expressed by the expectation values $\langle \hat{X} \rangle$ and $\langle \hat{Y} \rangle$ observed in separate measurements of the observables $\hat{X}$ and $\hat{Y}$. The statistics of the measurement outcomes is given by a joint probability $P_{\mathrm{exp.}}(x,y)$ over the four possible measurement outcomes $x,y=\pm1$. This probability distribution can also be characterized in terms of the averages of the measurement outcomes, $\langle x \rangle_{\mathrm{exp.}}$ and $\langle y \rangle_{\mathrm{exp.}}$. For a properly constructed measurement, these averages should be proportional to the expectation values $\langle \hat{X} \rangle$ and $\langle \hat{Y} \rangle$ of the input state, where random measurement errors result in a reduction of the proportionality by a constant factor. If these two factors are defined as the visibilities $V_x$ and $V_y$ of the joint measurement, the relations between the experimentally obtained averages of the measurement outcomes and the expectation values of the input state are given by
\begin{eqnarray}
\label{eq:V}
\langle x \rangle_{\mathrm{exp.}} &=& V_x \langle \hat{X} \rangle,
\nonumber \\
\langle y \rangle_{\mathrm{exp.}} &=& V_y \langle \hat{Y} \rangle.
\end{eqnarray} 
These relations strictly require that the experimental expectation values of the  measurement outcome $x$ depend only on the input expectation value $\langle \hat{X} \rangle$ of the quantum state, and the expectation values of the  measurement outcome $y$ depend only on the input expectation value $\langle \hat{Y} \rangle$ of the quantum state. In an experiment, it would be necessary to ensure that there is no dependence of the outcomes of $\hat{X}$ on $\hat{Y}$ or vice versa, since such relations would create artificial correlations between $x$ and $y$ in the measurement outcomes. 

The complete joint probability $P_{\mathrm{exp.}}(x,y)$ can be derived from the averages $\langle x \rangle_{\mathrm{exp.}}$ and $\langle y \rangle_{\mathrm{exp.}}$ and a third average for the product of the two outcomes, $\langle xy \rangle_{\mathrm{exp.}}$. This third average cannot be related directly to an expectation value of the qubit since it cannot be obtained in separate measurements of $\hat{X}$ and $\hat{Y}$. It thus represents a piece of experimental evidence about the relation between $\hat{X}$ and $\hat{Y}$ that can only be obtained in joint measurements \cite{Iin18,Kin15}. A complete characterization of the statistics of errors in joint quantum measurements therefore requires an additional description of how quantum theory determines this additional piece of information. 

For complementary properties of two level systems, it is possible to construct joint measurements that are only sensitive to the expectation values $\langle \hat{X} \rangle$ and $\langle \hat{Y} \rangle$ that span the equatorial plane of the Bloch vector. If these measurements satisfy the requirements of Eq.(\ref{eq:V}), there cannot be any systematic bias favoring either positive or negative values of the product $xy$ in the outcomes of the measurement. Therefore, the experimental average of the product $xy$ is exactly equal to zero for this class of joint measurements,
\begin{equation} 
\label{eq:xy}
\langle xy \rangle_{\mathrm{exp.}} = 0.
\end{equation}
This relation is indeed satisfied by a wide range of experimental realizations of joint measurements, such as the sequential measurements used to study Leggett-Garg inequality violations and measurement uncertainties \cite{Pry04,Bae08,Iin11,Gog11,Suz12,Roz12,Bae13,Rin14}. We should therefore consider Eq.(\ref{eq:xy}) as an experimentally confirmed property of uncertainty limited joint measurements of $\hat{X}$ and $\hat{Y}$. Conditions (\ref{eq:V}) and (\ref{eq:xy}) completely determine the four outcome probabilities $P_{\mathrm{exp.}}(x,y)$ of the joint measurement. The results read
\begin{eqnarray}
\label{eq:Pxy}
P_{\mathrm{exp.}}(+1,+1) &=& \frac{1}{4}\left(1 + V_x \langle \hat{X} \rangle + V_y \langle \hat{Y} \rangle \right)
\nonumber \\
P_{\mathrm{exp.}}(+1,-1) &=& \frac{1}{4}\left(1 + V_x \langle \hat{X} \rangle - V_y \langle \hat{Y} \rangle \right)
\nonumber \\
P_{\mathrm{exp.}}(-1,+1) &=& \frac{1}{4}\left(1 - V_x \langle \hat{X} \rangle + V_y \langle \hat{Y} \rangle \right)
\nonumber \\
P_{\mathrm{exp.}}(-1,-1) &=& \frac{1}{4}\left(1 - V_x \langle \hat{X} \rangle - V_y \langle \hat{Y} \rangle \right).
\end{eqnarray}
Because of condition (\ref{eq:xy}), these probabilities cannot describe an uncertainty free measurement. Specifically, the expectation value $\langle xy \rangle_{\mathrm{exp.}}$ would have to be equal to $+1$ for an uncertainty free measurement of an input with $x=+1$ and $y=+1$. If condition (\ref{eq:xy}) is imposed, the logical contradiction between the actual values and condition (\ref{eq:xy}) results in negative probabilities in Eq.(\ref{eq:Pxy}). It is therefore possible to identify the uncertainty limits required by Eq.(\ref{eq:xy}) quantitatively using the requirement of positivity for the probabilities in Eq.(\ref{eq:Pxy}). The result is a limit on visibilities and expectation values given by
\begin{equation}
|V_x \langle \hat{X} \rangle| + |V_y \langle \hat{Y} \rangle| \leq 1.
\end{equation}
The use of condition (\ref{eq:xy}) thus establishes a relation between the measurement uncertainties represented by $V_x$ and $V_y$ and the quantum state uncertainties represented by $\langle \hat{X} \rangle$ and $\langle \hat{Y} \rangle$. If we complete the description of the joint measurement by adding the uncertainty limit of $V_x$ and $V_y$, it is possible to derive the uncertainty limit for $\langle \hat{X} \rangle$ and $\langle \hat{Y} \rangle$ in quantum state preparation from the requirement that the statistics of the joint measurement must always be positive. 

For the class of measurements discussed here, the measurement uncertainties described by the visibilities are known to satisfy an uncertainty relation given by \cite{Iin11,Eng96}
\begin{equation}
\label{eq:MU}
V_x^2 + V_y^2 \leq 1.
\end{equation}  
Using this uncertainty limit, the requirement that the experimentally observed probabilities in Eq.(\ref{eq:Pxy}) must always be positive results in the condition
\begin{equation}
\langle \hat{X} \rangle^2 + \langle \hat{Y} \rangle^2 \leq 1
\end{equation}  
for all possible input states. Thus the curved surface of the Bloch sphere can be derived from the statistical properties of joint measurements given by condition (\ref{eq:xy}) and the measurement uncertainty in Eq.(\ref{eq:MU}). 

It is important to remember that the impossibility of uncertainty free joint measurements is an essential element of quantum theory \cite{Uol14}. In the derivation above, uncertainties emerge as a result of a single non-classical condition, Eq.(\ref{eq:xy}), which defines a necessary statistical relation between the measurement outcomes $x$ and $y$. Since this statistical relation does not depend on the input state, it is difficult to reconcile it with any measurement independent assignment of values to $x$ and $y$. This difficulty imposes fundamental limitations on quantum state statistics that are actually more restrictive than hidden variable theories. It is therefore interesting to ask what kind of restrictions we can obtain for quantum correlations that exceed the bounds of such hidden variable theories. Specifically, it should be interesting to analyze the specific bounds imposed by a pair of joint measurements on the quantum correlations that appear in Bell's inequalities.

\section{Observation of two qubit correlations by joint measurements}
\label{sec:joint}

If two local joint measurements are performed independently on a pair of qubits, the correlations between the two qubits will be observed in the joint measurement statistics $P_{\mathrm{exp.}}(x_A, y_A, x_B, y_B)$. In general, there are fifteen independent statistical moments that characterize this probability distribution over sixteen possible outcomes. To simplify the problem somewhat, the analysis can be limited to situations where all local expectation values are zero. The marginal probabilities for qubit $A$ and for qubit $B$ are then characterized by probabilities of $1/4$ for all four possible outcomes of $(x_A,y_A)$ or $(x_B,y_B)$. This assumption eliminates six of the fifteen statistical moments, leaving only nine moments of the distribution for the following analysis. These nine statistical moments can all be expressed as correlations between $(x_A,y_A,x_A y_A)$ and $(x_B,y_B,x_B y_B)$. However, Eq.(\ref{eq:xy}) implies that expectation values that involve either $x_A y_A$ or $x_B y_B$ will all be zero. If we represent an arbitrary contribution from system $i$ by $f_i$, the rule for correlations between the outcomes of joint measurements reads
\begin{eqnarray}
\label{eq:corr0}
\langle x_A y_A \; f_B \rangle_{\mathrm{exp.}} &=& 0, 
\nonumber \\ 
\langle f_A \; x_B y_B \rangle_{\mathrm{exp.}} &=& 0.
\end{eqnarray}
Note that $f_i$ can stand for $x_i$, $y_i$ or the product $x_i y_i$. This means that the first line of Eq.(\ref{eq:corr0}) represents three additional conditions, and the second line provides another two, with the third condition being equal to the third condition of the first line. In total, these are five more conditions that determine the joint measurement statistics for the two independently performed local measurements of the two qubits. The remaining four statistical moments are determined by the correlations between $(x_A,y_A)$ and $(x_B,y_B)$, which can be related to the input expectation values between $(\hat{X}_A,\hat{Y}_A)$ and $(\hat{X}_B,\hat{Y}_B)$ using the visibilities of local measurements as defined by Eq.(\ref{eq:V}),
\begin{eqnarray}
\label{eq:2QbitV}
\langle x_A x_B \rangle_{\mathrm{exp.}} &=& V_x(A) V_x(B) \langle \hat{X}_A \otimes \hat{X}_B \rangle,
\nonumber \\
\langle x_A y_B \rangle_{\mathrm{exp.}} &=& V_x(A) V_y(B) \langle \hat{X}_A \otimes \hat{Y}_B \rangle,
\nonumber \\
\langle y_A x_B \rangle_{\mathrm{exp.}} &=& V_y(A) V_x(B) \langle \hat{Y}_A \otimes \hat{X}_B \rangle,
\nonumber \\
\langle y_A y_B \rangle_{\mathrm{exp.}} &=& V_y(A) V_y(B) \langle \hat{Y}_A \otimes \hat{Y}_B \rangle.
\end{eqnarray} 
The probabilities $P_{\mathrm{exp.}}$ for the sixteen outcomes of the two joint measurements $(x_A, y_A, x_B, y_B)$ thus depend only on the four correlations between local observables that can also be observed in separate measurements of $\hat{X}_A$ or $\hat{Y}_A$ and  $\hat{X}_B$ or $\hat{Y}_B$. All of the other 11 experimentally observed expectation values are exactly zero. 

It is in principle a straightforward matter to derive explicit expressions for the experimentally observable probabilities by inverting the relations expressing the 15 experimental averages and the total probability of one as linear combinations of the joint probabilities. Since only four of the experimental averages are non-zero, the resulting expressions for the probabilities take a rather simple form. Although it would be possible to give a complete list of the 16 expressions, it might be sufficient to show one explicit example for which the positivity limit is particularly easy to see,
\begin{eqnarray}
\label{eq:jointP}
\lefteqn{P_{\mathrm{exp.}}(+1,+1,-1,-1) =} 
\nonumber \\ && \hspace{0.5cm}
\frac{1}{16} (1 - V_x(A) V_x(B) \langle \hat{X}_A \otimes \hat{X}_B \rangle - V_x(A) V_y(B) \langle \hat{X}_A \otimes \hat{Y}_B \rangle
\nonumber \\ && \hspace{1.5cm}
- V_y(A) V_x(B) \langle \hat{Y}_A \otimes \hat{X}_B \rangle
- V_y(A) V_y(B) \langle \hat{Y}_A \otimes \hat{Y}_B \rangle) \geq 0.
\end{eqnarray}
For other values of $(x_A, y_A, x_B, y_B)$, the signs in front of each correlation are determined by the corresponding products of $(x_A, y_A)$ and $(x_B, y_B)$. As will be shown in the next section, these changes in sign before the visibilities can be used to identify a single expression for the requirement that all 16 experimental probabilities must always be positive. 

Eq. (\ref{eq:jointP}) shows that the experimentally observed probability distribution of a joint measurement of $(x_A, y_A, X_B, y_B)$ originates from the quantum state statistics observed in separate projective measurements on $\hat{X}_A$ or $\hat{Y}_A$ and  $\hat{X}_B$ or $\hat{Y}_B$. It is therefore possible to predict the measurement probabilities of the joint measurement from the results of separate measurements of the correlations $\langle \hat{X}_A \otimes \hat{X}_B \rangle$, $\langle \hat{X}_A \otimes \hat{Y}_B \rangle$, $\langle \hat{Y}_A \otimes \hat{X}_B \rangle$ and $\langle \hat{Y}_A \otimes \hat{Y}_B \rangle$. Since experimental measurement probabilities cannot be negative, Eq.(\ref{eq:jointP}) imposes a fundamental limit on the possible values of these correlations. In the following, these limits will be determined based on the known uncertainty limits for the visibilities of the measurements given by $V_x(A), V_y(A)$ and $V_x(B),V_y(B)$.

\section{Analysis of the statistical bounds for non-local correlations}

Since the probabilities of all possible experimental outcomes must be positive, it is possible to identify statistical bounds that relate local (single qubit) measurement uncertainties to non-local (two qubit) expectation values. For the probability in Eq.(\ref{eq:jointP}), this statistical bound is given by
\begin{eqnarray}
\label{eq:bound}
V_x(A) V_x(B) \langle \hat{X}_A \otimes \hat{X}_B \rangle &+& V_x(A) V_y(B) \langle \hat{X}_A \otimes \hat{Y}_B\rangle \nonumber \\
+ V_y(A) V_x(B) \langle \hat{Y}_A \otimes \hat{X}_B \rangle
&+& V_y(A) V_y(B) \langle \hat{Y}_A \otimes \hat{Y}_B \rangle \leq 1.
\end{eqnarray}
All other bounds can be obtained by flipping the signs in front of each visibility $V_m(S)$ to find the bound for an outcome with opposite sign in $m_S$. 

The bound given by Eq.(\ref{eq:bound}) depends on the visibilities $V_m(S)$, and these visibilities are themselves bounded by the uncertainty relation given in Eq.(\ref{eq:MU}). Since experimental probabilities must remain positive for all possible measurement uncertainties, the bounds for the correlations between qubit $A$ and qubit $B$ must ensure that no permitted combination of visibilities violates Eq.(\ref{eq:bound}) or any of the bounds obtained for the other measurement outcomes. Fortunately, it is possible to summarize the bounds by noting that the maximal visibilities satisfying Eq.(\ref{eq:MU}) can be characterized by trigonometric functions, so that the balance between the visibility $V_x(A)$ and the visibility $V_y(A)$ is described by an angle of $\alpha$, while the balance between the visibility of $V_x(B)$ and the visibility $V_y(B)$ is described by an angle of $\beta$. The constraint that visibilities need to be positive would seem to indicate that the angles $\alpha$ and $\beta$ should be limited to an interval between zero and $\pi/2$. However, this restriction can be lifted to formulate a single condition for all possible combination of measurement outcomes. Specifically, it is possible to define the visibilities as a product of trigonometric functions and the values of $(x_A, y_A, x_B, y_B)$ to which the condition of positive probability is to be applied,  
\begin{eqnarray}
\label{eq:angles}
V_x(A) &=& x_A \cos(\alpha)
\nonumber \\
V_y(A) &=& y_A \sin(\alpha)
\nonumber \\
V_x(B) &=& - x_B \cos(\beta)
\nonumber \\
V_y(B) &=& - y_B \sin(\beta).
\end{eqnarray}
For any combination of $\alpha$ and $\beta$, these visibilities are positive for one specific combination of $(x_A, y_A, x_B, y_B)$. When these visibilities are inserted into the formula for the joint probability of this specific combination of measurement outcomes, the values of the measurement outcomes cancel out, resulting in a single inequality that must be satisfied for all possible values of $\alpha$ and $\beta$,
\begin{eqnarray}
\label{eq:alphabet}
&&\cos(\alpha+\beta) \left(\langle \hat{X}_A \otimes \hat{X}_B \rangle - \langle \hat{Y}_A \otimes \hat{Y}_B \rangle\right)
\nonumber \\
 &+&  \sin(\alpha+\beta) \left(\langle \hat{X}_A \otimes \hat{Y}_B \rangle + \langle \hat{Y}_A \otimes \hat{X}_B \rangle\right)
\nonumber \\
&+& \cos(\alpha-\beta) \left(\langle \hat{X}_A \otimes \hat{X}_B \rangle + \langle \hat{Y}_A \otimes \hat{Y}_B \rangle\right) 
\nonumber \\
&-&  \sin(\alpha-\beta) \left(\langle \hat{X}_A \otimes \hat{Y}_B \rangle - \langle \hat{Y}_A \otimes \hat{X}_B \rangle\right) \leq 2.
\end{eqnarray}
In this inequality, $\alpha$ and $\beta$ represent both the uncertainty trade-off at the uncertainty limit given by Eq.(\ref{eq:bound}) and the selection of a specific measurement outcome for which the bound keeps the probabilities positive. It is then possible to find the bound for the quantum correlations of the two qubit state by requiring that Eq.(\ref{eq:alphabet}) must be satisfied for all possible values of $\alpha$ and $\beta$. 

Since the sums and the differences of $\alpha$ and $\beta$ can be varied independently, it is easy to identify the bound that applies to the correlations. Specifically, the optimal choice of visibilities results in contributions that correspond to the lengths of two dimensional vectors in the Bloch space of multi-qubit expectation values,
\begin{eqnarray}
\label{eq:tightB}
&& \sqrt{\left(\langle \hat{X}_A \otimes \hat{X}_B \rangle - \langle \hat{Y}_A \otimes \hat{Y}_B \rangle\right)^2 + \left(\langle \hat{X}_A \otimes \hat{Y}_B \rangle + \langle \hat{Y}_A \otimes \hat{X}_B \rangle\right)^2}
\nonumber \\ &+&
\sqrt{\left(\langle \hat{X}_A \otimes \hat{X}_B \rangle + \langle \hat{Y}_A \otimes \hat{Y}_B \rangle\right)^2 +  \left(\langle \hat{X}_A \otimes \hat{Y}_B \rangle - \langle \hat{Y}_A \otimes \hat{X}_B \rangle\right)^2} \leq 2.
\end{eqnarray}
This bound is the central result of the present paper. It describes a tight statistical limit for the four non-local correlations between a pair of qubits that also appear in Bell's inequalities. However, the present bound is based on the necessary positivity of experimentally observable probabilities, and not on the hypothetical possibility of assigning hidden variables to unobserved quantities. Specifically, the bound given by Eq.(\ref{eq:tightB}) is necessary because quantum mechanics allows for joint local measurements that would result in experimentally observable negative probabilities if this bound was exceeded. The possibility of uncertainty limited joint measurements thus explains why  no quantum state can ever exceed the bound given by Eq.(\ref{eq:tightB}). Since the bound is a fundamental limit of the quantum correlations between two qubits, it is closely related to the Cirel'son bound \cite{Cir80}, which describes the maximal violation of Bell's inequalities permitted by quantum mechanics. In fact, the Cirel'son bound is already included in the bound given by Eq.(\ref{eq:tightB}), as will be shown in the following.

To obtain a better intuitive understanding of the bound in Eq.(\ref{eq:tightB}), it may be useful to derive bounds that are less tight but easier to interpret. The most obvious simplification is to focus on only one of the two square roots on the right hand side of the inequality, which allows us to remove the square root to obtain
\begin{equation}
\label{eq:halfB}
\left(\langle \hat{X}_A \otimes \hat{X}_B \rangle - \langle \hat{Y}_A \otimes \hat{Y}_B \rangle\right)^2 + \left(\langle \hat{X}_A \otimes \hat{Y}_B \rangle + \langle \hat{Y}_A \otimes \hat{X}_B \rangle\right)^2 \leq 4.
\end{equation}
This simplification of the bound already brings us closer to the term on the left hand side of a Bell's inequality. Specifically, the Bell's inequality defines a bound for the sum of the two terms that are squared on the left hand side of Eq.(\ref{eq:halfB}). Using the basic mathematical properties of squares, it is possible to identify the bound that applies only to the sum of the terms, which can be saturated when the difference of the two terms is exactly equal to zero. This bound for the left hand side of the Bell's inequality reads
\begin{equation}
\label{eq:Cirel}
\langle \hat{X}_A \otimes \hat{X}_B \rangle + \langle \hat{X}_A \otimes \hat{Y}_B \rangle + \langle \hat{Y}_A \otimes \hat{X}_B \rangle - \langle \hat{Y}_A \otimes \hat{Y}_B \rangle \leq 2 \sqrt{2},
\end{equation}
which is equal to the Cirel'son bound that described the maximal violation of Bell's inequalities allowed by quantum mechanics \cite{Cir80}.

The relation between the Cirel'son bound and joint measurability has been identified previously by Banik and coworkers \cite{Ban13}. However, the joint measurement considered in that work was a non-local measurement of the two correlations that are squared on the right hand side of Eq.(\ref{eq:halfB}). Effectively, Banik et al. considered the incompatibility between measurements of $\hat{X}_A\hat{X}_B-\hat{Y}_A\hat{Y}_B$ and $\hat{X}_A\hat{Y}_B+\hat{X}_A\hat{Y}_B$ in the two dimensional subspace of the two qubit Hilbert space where these two correlations have eigenvalues of $\pm 2$. It is therefore an important result that the bound can also be derived from the uncertainty limits of local measurements. The present analysis shows that the Cirel'son bound can be explained by the measurement uncertainties of two local measurements, where each measurement outcome assigns individual values to each of the four local observables. The combined effects of non-local quantum state statistics and local measurement uncertainties can then be observed in the joint probabilities of the 16 possible outcomes. In particular, achievement of the tight bound given by Eq. (\ref{eq:tightB}) means that a specific joint probability drops to zero at this bound, and any further increase in the correlations responsible for the low probability of that outcome would reduce this probability to a negative value. Oppositely, any increase in measurement visibility would likewise result in a negative value of the probability. Since the visibilities are limited by local uncertainty relations, the present result indicates that the failure of local realism associated with the violation of Bell's inequalities originates from a local failure of realism associated with measurement uncertainties. As explained in Sec. \ref{Sec:measure}, local uncertainty bounds are also needed to keep experimental probabilities positive for local input states. The Cirel'son bound can thus be explained a non-local consequence of this entirely local relation between measurement uncertainties and the statistics of quantum states.

\section{Saturation of the bound}

Before taking a closer look at the statistics observed at the bound, it may be good to relate the bound to the quantum states that saturate it. In particular, it is important to demonstrate that the bound is tight and cannot be improved upon by the addition of more terms. To do so, it is necessary to identify the density matrix elements that correspond to the correlations in Eq.(\ref{eq:tightB}). Using the conventional definition of the computational basis $\{\mid 0 \rangle, \mid 1 \rangle\}$ so that $\hat{X} \mid 0 \rangle = \mid 1 \rangle$ and $\hat{Y} \mid 0 \rangle = i \mid 1 \rangle$, the correlations can be identified with specific off-diagonal elements of the density matrix using
\begin{eqnarray}
(\hat{X}_A \otimes \hat{X}_B - \hat{Y}_A \otimes \hat{Y}_B) -i (\hat{X}_A \otimes \hat{Y}_B + \hat{Y}_A \otimes \hat{X}_B ) &=& 4 \mid 11 \rangle \langle 00 \mid
\nonumber \\
(\hat{X}_A \otimes \hat{X}_B + \hat{Y}_A \otimes \hat{Y}_B) -i (\hat{X}_A \otimes \hat{Y}_B - \hat{Y}_A \otimes \hat{X}_B ) &=& 4 \mid 01 \rangle \langle 10 \mid.
\end{eqnarray}
The bound given by Eq.(\ref{eq:tightB}) is therefore equal to the mathematical limit on two coherences in separate subspaces of the density matrix,
\begin{equation}
|\langle 00 \mid \hat{\rho} \mid 11 \rangle| + |\langle 10 \mid \hat{\rho} \mid 01 \rangle| \leq \frac{1}{2}.
\end{equation}
This bound is saturated by any equal superpositions of $\mid 00 \rangle$ and $\mid 11 \rangle$, by any equal superpositions of $\mid 10 \rangle$ and $\mid 01 \rangle$, and by any statistical mixture of the two.

The simplified bound of Eq.(\ref{eq:halfB}) is the bound in the subspace of $\mid 00 \rangle$ and $\mid 11 \rangle$, saturated by any state of the form
\begin{equation}
\label{eq:psisat}
\mid \psi_{\mathrm{sat.}} \rangle = \frac{1}{\sqrt{2}} \left(\mid 00 \rangle + \exp(i \phi) \mid 11 \rangle\right),
\end{equation}
with correlations of 
\begin{eqnarray}
\label{eq:satcor}
\langle \hat{X}_A \otimes \hat{X}_B \rangle = - \langle \hat{Y}_A \otimes \hat{Y}_B \rangle = \cos(\phi), \nonumber \\
\langle \hat{X}_A \otimes \hat{Y}_B \rangle = \langle \hat{Y}_A \otimes \hat{X}_B \rangle = \sin(\phi).
\end{eqnarray}
The bound thus identifies a maximal two qubit coherence. Note that the Cirel'son bound given by Eq.(\ref{eq:Cirel}) is only saturated for $\phi=\pi/4$, demonstrating that the bounds derived here are significantly tighter than the Cirel'son bound itself. 

\section{Experimental statistics at the bound}

The upper bound on non-local correlations between two separate qubits derived above originates from the requirement that the experimentally observable measurement probabilities of joint measurements must remain positive for all physical input states. This means that the bounds can be tested experimentally by showing that the saturation of the bound corresponds to a probability of zero in a corresponding joint measurement. It is interesting to relate this observation to the previous result that the Cirel'son bound can be explained by the uncertainty limit for conditional statistics observed in correlated qubits \cite{Opp10}. For a quantum state at the bound, there exists a combination of joint measurements so that at least one of the outcome probabilities is zero. If the measurement of system A is treated as part of a conditional quantum state preparation in B, the conditional probabilities in system B describe the joint measurement of a single qubit with one of the four joint probabilities at zero. From the discussion in Sec. \ref{sec:joint}, it is clear that the joint probability of zero marks the uncertainty limit for $\hat{X}_B$ and $\hat{Y}_B$ in system B. Therefore the present results may also serve as a generalization of the conditional uncertainty bound introduced by Oppenheim and Wehner \cite{Opp10} to a wider variety of possible remote state preparations. The interpretation of the bound as a conditional uncertainty bound may also help to explain why Bell's inequality violations are possible even though the local uncertainty bounds are more restrictive than hidden variable theories. Since the joint measurement in A used to select the conditional state in B is also uncertainty limited, the conditional reduction of uncertainties in B is lower than the reduction caused by a hypothetical discovery of the hidden variables that determine the exact values of $\hat{X}_A$ and $\hat{Y}_A$ simultaneously. Measurement uncertainties thus identify a gap between the tighter bounds of hidden local realism and the more permissive bounds of observable local reality.

Finally, it may also be helpful to look at the experimentally observed correlations in more detail. Banik and coworkers derived the Cirel'son bound by arguing that the experimental probabilities of a collective (non-local) measurement were limited by Bell's inequalities and demonstrating that the minimal factor by which the initial correlations were reduced was $\sqrt{2}$ \cite{Ban13}. The present argument is quite different, since the positivity bounds of the individual outcome probabilities are more detailed and precise than the collective bounds imposed by Bell's inequalities. We can therefore expect that the experimental probabilities at the bound do not saturate Bell's inequalities. For the state in Eq.(\ref{eq:psisat}), the experimentally observed correlation is
\begin{eqnarray}
\langle x_A x_B \rangle_{\mathrm{exp.}} &+&
\langle y_A x_B \rangle_{\mathrm{exp.}} +
\langle x_A y_B \rangle_{\mathrm{exp.}} -
\langle y_A y_B \rangle_{\mathrm{exp.}} 
\nonumber \\ &&
= \cos(\alpha-\beta) \cos (\phi) + \sin(\alpha+\beta) \sin (\phi),
\end{eqnarray}
where $0<\alpha<\pi/2$ and $0<\beta<\pi/2$ describe the visibilities of the two local measurements according to Eq.(\ref{eq:angles}). The maximal value of this correlation is $\sqrt{2}$, obtained for $\alpha=\beta=\pi/4$, which describes measurements with equal visibilities of $V_x=V_y=1/\sqrt{2}$ in both systems. This actually seems to be the highest experimental value for this correlation in any joint measurement, indicating that the Bell bound of two cannot be achieved by the experimental statistics of local joint measurements. In fact, the experimental correlation is even lower for measurements that result in an experimental probability of zero. In this case, the measurement visibilities must satisfy $\alpha+\beta=\phi$ and the maximal Bell`s inequality correlation is obtained for $\alpha=\beta=\phi/2$. In the presence of an experimental outcome probability of zero, the correlation for the state in Eq.(\ref{eq:psisat}) is given by
\begin{eqnarray}
\langle x_A x_B \rangle_{\mathrm{exp.}} +
\langle y_A x_B \rangle_{\mathrm{exp.}} &+&
\langle x_A y_B \rangle_{\mathrm{exp.}} -
\langle y_A y_B \rangle_{\mathrm{exp.}} 
\nonumber \\ && \hspace{0.8cm}
= 1+\cos (\phi)- \cos^2 (\phi).
\end{eqnarray}
The maximal experimentally observable value of the Bell's inequality correlation in the presence of a measurement outcome with probability zero is $1.25$, achieved at $\cos(\phi)=1/2$. Interestingly, the corresponding state does not saturate the Cirel'son bound, since the Bell's inequality violation of the correlations shown in Eq.(\ref{eq:satcor}) is only $(1+\sqrt{3})\approx 2.73$ for this value of $\phi$. This result highlights the difference between the bound described by Bell's inequalities and the actual quantum mechanical bound required to obtain only positive experimental probabilities. 

The problem with Bell's inequalities seems to be that they do not really identify any relevant quantum mechanical limit at all. Local uncertainty limits indicating entanglement are well below the Bell's inequality bound and the Cirel'son bound is well above it. Even the experimentally observable probabilities of joint measurements fail to come close to the bound of local realism. The reason for the latter failure to achieve the Bell's inequality bound is that the positivity of experimentally observable probabilities must accommodate not just the Bell's inequality sum of correlations, but also a number of other conditions, as summarized by the comprehensive bound on non-local correlations given by Eq.(\ref{eq:tightB}). It may therefore be justified to conclude that the statistical bounds imposed by local joint measurements are more fundamental than the quantum non-locality evidenced by a violation of Bell's inequalities. 

\section{Conclusions}
 
Uncertainty bounds on joint measurability are a necessary condition for the observation of non-classical correlations between non-commuting observables \cite{Wol09,Qui14,Uol14}. As shown above, it is possible to trace the non-classical relation between complementary physical properties of qubit systems to a simple rule for the probabilities of measurement outcomes observed in uncertainty limited joint measurements of these properties, which is that the average product of the two measurement outcomes of $\pm 1$ must always be zero (Eq.(\ref{eq:xy})). When combined with the uncertainty limit of Eq.(\ref{eq:MU}), this simple rule not only determines the uncertainty limit of the qubit Bloch sphere, but also imposes collective statistical limits on pairs of qubits that relate specifically to the maximal values of two qubit correlations. The limit given by Eq.({\ref{eq:tightB}) explains the statistical signature of maximal two qubit coherence as a consequence of the uncertainty limits of local joint measurements. This limit naturally includes the Cirel'son bound as only one component of the more general limit on two qubit correlations. It is therefore possible to conclude that the uncertainty limits of joint quantum measurements imposed by the experimentally observable correlations between non-commuting observables are more fundamental than the associated limit imposed on the violation of Bell's inequalities by the Cirel'son bound.  

The present result indicates that the limitations of non-local quantum correlations are not related to any additional constraints on non-local hidden variable models, but originate from local features of the statistical relations between non-commuting observables that appear directly in the experimental data of joint measurements of the local systems. Specifically, joint measurements give direct access to the non-classical relations between physical properties that seem to be hidden by the statistical uncertainties described by the Hilbert space formalism. These relations strongly suggest that there is no joint reality of these properties, independent of whether the state is separable or entangled. It is therefore entirely possible - and perhaps even likely \cite{Cab10,Liu16}) - that the violation of Bell's inequalities is merely a natural consequence of the local structure of quantum statistics. 

\section*{Acknowledgments}
This work was supported by JSPS KAKENHI Grant Number 26220712.


\section*{References}
\begin{thebibliography}{10}


\bibitem{Pry04}
G. J. Pryde, J.L. O’Brien, A. G. White, S. D. Bartlett, and T. C. Ralph, ``Measuring a photonic qubit without destroying it,'' Phys. Rev. Lett. {\bf 92}, 190402 (2004).

\bibitem{Bae08}
S.-Y. Baek, Y. W. Cheong, and Y.-H. Kim, ``Minimum-disturbance measurement without postselection,'' Phys. Rev. A {\bf 77}, 060308(R) (2008).

\bibitem{Yok09} 
K. Yokota, T. Tamamoto, M. Koashi, and N. Imoto, ``Direct observation of Hardy's paradox by joint weak measurement with an entangled photon pair,'' New J. Phys. {\bf 11}, 033011 (2009). 

\bibitem{Ste09} 
J. S. Lundeen and A. M. Steinberg, ``Experimental Joint Weak Measurement on a Photon Pair as a Probe of Hardys Paradox,'' Phys. Rev. Lett. {\bf 102}, 020404 (2009).

\bibitem{Iin11} 
M. Iinuma, Y. suzuki, G. Taguchi, Y. Kadoya and H. F. Hofmann, ``Weak measurement of photon polarization by back-action-induced path interference,'' New J. Phys. {\bf 13}, 033041 (2011).

\bibitem{Gog11}
M. E. Goggin, M. P. Almeida, M. Barbieri, B. P. Lanyon, J. L. O'Brien, A. G. White, and G.J. Pryde,
``Violation of the Leggett–Garg inequality with weak measurements of photons,'' Proc. Natl. Acad. Sci. U. S. A. {\bf 108} 1256 (2011).

\bibitem{Suz12} 
Y. Suzuki, M. Iinuma, and H. F. Hofmann, ``Violation of Leggett–Garg inequalities in quantum measurements with variable resolution and back-action,'' New J. Phys. {\bf 14}, 103022 (2012).

\bibitem{Roz12}
L. A. Rozema, A. Darabi, D. H. Mahler, A. Hayat, Y. Soudagar,
and A. M. Steinberg, ``Violation of Heisenberg’s Measurement-Disturbance Relationship by Weak Measurements,'' Phys. Rev. Lett. {\bf 109}, 100404 (2012). 

\bibitem{Tan13}
J.-S. Tang, Y.-L. Li, C.-F. Li, and G.-C. Guo, ``Revisiting Bohr's principle of complementarity with a quantum device,'' Rev. A {\bf 88}, 014103 (2013).

\bibitem{Bae13}
S.-Y. Baek, F. Kaneda, M. Ozawa, and K. Edamatsu, ``Experimental violation and reformulation of the Heisenberg's error-disturbance uncertainty relation,'' {\it Sci. Rep.} {\bf 3}, 2221 (2013).

\bibitem{Boy13} 
J.Z. Salvail, M. Agnew, S. J. Bolduc, J. Leach and R. W. Boyd, ``Full characterization of polarization states of light via direct measurement,'' Nat. Photon. {\bf 7}, 316-321 (2013).

\bibitem{Rin14}
M. Ringbauer, D. N. Biggerstaff, M. A. Broome, A. Fedrizzi, C. Branciard, and A. G. White, ``Experimental Joint Quantum Measurements with Minimum Uncertainty,''  Phys. Rev. Lett. {\bf 112}, 020401 (2014).

\bibitem{Kan14}
F. Kaneda, S.-Y. Baek, M. Ozawa, and K. Edamatsu, ``Experimental Test of Error-Disturbance Uncertainty Relations by Weak Measurement, ''  Phys. Rev. Lett. {\bf 112}, 020402 (2014).

\bibitem{Den14}
T. Denkmayr, H. Geppert, S. Sponar, H. Lemmel, A. Matzkin, J. Tollaksen, and Y. Hasegawa, ``Observation of a quantum Cheshire Cat in a matter-wave interferometer experiment,'' Nat. Commun. {\bf 5}, 4492 (2014).

\bibitem{Bam14} 
C. Bamber and J. S. Lundeen, ``Observing Dirac's Classical Phase Space Analog to the Quantum State,''  Phys. Rev. Lett. {\bf 112}, 070405 (2014). 

\bibitem{Val16} 
G. Vallone and D. Dequal, ``Strong measurements give a better direct measurement of the quantum wave function,'' Phys. Rev. Lett. {\bf 116}, 040502 (2016).

\bibitem{Pia16}
F. Piacentini, A. Avella, M. P. Levi, M. Gramegna, G. Brida, I. P. Degiovanni, E. Cohen, R. Lussana, F. Villa, A. Tosi, F. Zappa, and M. Genovese, ``Measuring Incompatible Observables by Exploiting Sequential Weak Values,'' Phys. Rev. Lett. {\bf 117}, 170402 (2016).

\bibitem{Iin16}
M. Iinuma, Y. Suzuki, T. Nii, R. Kinoshita, and H. F. Hofmann, ``Experimental evaluation of nonclassical correlations between measurement outcomes and target observable in a quantum measurement,'' Phys. Rev. A {\bf 93}, 032104 (2016).

\bibitem{Suz16}
Y. Suzuki, M. Iinuma, and H. F. Hofmann, ``Observation of non-classical correlations in sequential measurements of photon polarization,'' New J. Phys. {\bf 18}, 103045 (2016). 

\bibitem{Pia17}
F. Piacentini, A. Avella, E. Rebufello, R. Lussana, F. Villa, A. Tosi, M. Gramegna, G. Brida, E. Cohen, L. Vaidman, I. P. Degiovanni, and M. Genovese, ``Determining the quantum expectation value by measuring a single photon,'' Nat. Phys. {\bf 13}, 1191 (2017).

\bibitem{The17}
G. S. Thekkadath, R. Y. Saaltink, L. Giner, and J. S. Lundeen, ``Determining Complementary Properties with Quantum Clones,'' Phys. Rev. Lett. {\bf 119}, 050405 (2017).

\bibitem{Iin18}
M. Iinuma, M. Nakano, H. F. Hofmann, and Y. Suzuki, ``Experimental evaluation of the non-classical relation between measurement errors using entangled photon pairs as a probe,'' Phys. Rev. A {\bf 98}, 062109 (2018).


\bibitem{Eng96}
B. G. Englert, ``Fringe Visibility and Which-Way Information: An Inequality,''
Phys. Rev. Lett. {\bf 77}, 2154-7 (1996).

\bibitem{And05}
E. Andersson, S. M. Barnett, and A. Aspect, ``Joint measurements of spin, operational locality, and uncertainty,'' Phys. Rev. A {\bf 72}, 042104 (2005).

\bibitem{Hei08}
T. Heinosaari, D. Reitzner, and P. Stano, ``Notes on Joint Measurability of Quantum Observables,''  Found. Phys. {\bf 38}, 1133-1147 (2008).

\bibitem{Sta08}
P. Stano, D. Reitzner, and T. Heinosaari, ``Coexistence of qubit effects,'' Phys. Rev. A {\bf 78}, 012315 (2008).

\bibitem{Yu10}
S. Yu, N.-L. Liu, L. Li, and C. H. Oh, ``Joint measurement of two unsharp observables of a qubit,'' Phys. Rev. A {\bf 81}, 062116 (2010).

\bibitem{Hei10}
T. Heinosaari, M. A. Jivulescu, D. Reitzner, and M. Ziman, ``Approximating incompatible von Neumann measurements simultaneously,''  Phys. Rev. A {\bf 82}, 032328 (2010).

\bibitem{Car11}
C. Carmeli, T. Heinosaari, and A. Toigo, ``Sequential measurements of conjugate observables,'' J. Phys. A: Math. Gen.  {\bf 44}, 285304 (2011). 

\bibitem{Lor11}
A. Di Lorenzo, ``Strong correspondence principle for joint measurement of conjugate observables,'' Phys. Rev. A {\bf 83}, 042104 (2011).

\bibitem{Car12}
C. Carmeli, T. Heinosaari, and A. Toigo, ``Informationally complete joint measurements on finite quantum systems,'' Phys. Rev. A {\bf 85}, 012109 (2012).

\bibitem{Bus13}
P. Busch, T. Heinosaari, J. Schultz, and N. Stevens, ``Comparing the degrees of incompatibility inherent in probabilistic physical theories,'' EPL {\bf 103}, 10002 (2013).

\bibitem{Hei13}
T. Heinosaari, ``A simple sufficient condition for the coexistence of quantum effects,'' J. Phys. A: Math. Gen.  {\bf 46}, 152002 (2013). 

\bibitem{Hof14}
H. F. Hofmann, ``Sequential measurements of non-commuting observables with quantum controlled interactions,'' New J. Phys. {\bf 16}, 063056 (2014). 

\bibitem{Kun14}
R. Kunjwal, C. Heunen, and T. Fritz, ``Quantum realization of arbitrary joint measurability structures,'' Phys. Rev. A {\bf 89}, 052126 (2014). 

\bibitem{Kin15}
S. Kino, T. Nii, and H. F. Hofmann, ``Characterization of measurement uncertainties using the correlations between local outcomes obtained from maximally entangled pairs,'' Phys. Rev. A {\bf 92}, 042113 (2015). 

\bibitem{Kaw18}
R. Kawakubo and T. Koike, ``Disturbance by optimal discrimination,'' Phys. Rev. A {\bf 97}, 032102 (2018).


\bibitem{Bell}
J. S. Bell, ``On the Einstein Podolsky Rosen paradox,'' Physics {\bf 1}, 195-200 (1964).

\bibitem{Gen05}
M. Genovese, ``Research on hidden variable theories: A review of recent progresses,'' Phys. Rep. {\bf 413}, 319 (2005).

\bibitem{Car18}
A. Carmi and E. Cohen, ``On the Significance of the Quantum Mechanical Covariance Matrix,'' Entropy {\bf 2018}, 20, 500 (2018).

\bibitem{Wol09}
M. M. Wolf, D. Perez-Garcia, and C. Fernandez, ``Measurements Incompatible in Quantum Theory Cannot Be Measured Jointly in Any Other No-Signaling Theory,'' Phys. Rev. Lett. {\bf 103}, 230402 (2009).

\bibitem{Qui14}
M. T. Quintino, T. Vertesi, and N. Brunner, ``Joint Measurability, Einstein-Podolsky-Rosen Steering, and Bell Nonlocality,'' Phys. Rev. Lett. {\bf 113}, 160402 (2014).

\bibitem{Uol14}
R. Uola, T. Moroder, and O. G\"uhne, ``Joint Measurability of Generalized Measurements Implies Classicality,'' Phys. Rev. Lett. {\bf 113}, 160403 (2014). 


\bibitem{Ste14}
N. Stevens and P. Busch, ``Steering, incompatibility, and Bell-inequality violations in a class of probabilistic theories,'' Phys. Rev. A {\bf 89}, 022123 (2014).

\bibitem{Kar15}
H. S. Karthik, A.~R.~Usha  Devi, and A. K. Rajagopal, ``Joint measurability, steering, and entropic uncertainty,'' Phys. Rev. A {\bf 91}, 012115 (2015).

\bibitem{Qui16}
M. T. Quintino, J. Bowles, F. Hirsch, and N. Brunner, ``Incompatible quantum measurements admitting a local-hidden-variable model,'' Phys. Rev. A {\bf 93}, 052115 (2016).
 
\bibitem{Hir18}
F. Hirsch, M. T. Quintino, and N. Brunner, ``Quantum measurement incompatibility does not imply Bell nonlocality,'' Phys. Rev. A {\bf 97}, 012129 (2018).


\bibitem{Cir80}
B. S. Cirel'son, ``Quantum generalizations of Bell's inequality,'' Lett. Math. Phys. {\bf 4}, 93-100 (1980).

\bibitem{Opp10}
J. Oppenheim and S. Wehner, ``The Uncertainty Principle Determines the Nonlocality of Quantum Mechanics,'' Science {\bf 330}, 1072 (2010).

\bibitem{Car19}
A more general formulation of the relation between quantum state uncertainties and nonlocality recently appeared in A. Carmi, E. Cohen, ``Relativistic independence bounds nonlocality,'' Sci. Adv. {\bf 5}, eaav8370 (2019). 

\bibitem{Ban13}
M. Banik, Md.~Rajjak Gazi, S. Ghosh and G. Kar, ``Degree of complementarity determines the nonlocality in quantum mechanics,'' Phys. Rev. A {\bf 87}, 052125 (2013).

\bibitem{Cab10}
A. Cabello, ``Proposal for Revealing Quantum Nonlocality via Local Contextuality,'' Phys. Rev. Lett. {\bf 104}, 220401 (2010). 

\bibitem{Liu16}
B.-H. Liu, X.-M. Hu, J.-S. Chen, Y.-F. Huang, Y.-J. Han, C.-F. Li, G.-C. Guo, and A. Cabello, ``Nonlocality from Local Contextuality,'' Phys. Rev. Lett. {\bf 117}, 220402 (2016).

\end{thebibliography}
\end{document}